\documentclass[pdflatex,sn-mathphys-num]{sn-jnl}
\usepackage{fix-cm}  % allows scalable Computer Modern fonts
\usepackage{scalefnt}
\usepackage{mathrsfs}
\usepackage{graphicx}%
\usepackage{multirow}%
\usepackage{amsmath,amssymb,amsfonts}%
\usepackage{amsthm}%
\usepackage{mathrsfs}%
\usepackage[title]{appendix}%
\usepackage{xcolor}%
\usepackage{textcomp}%
\usepackage{manyfoot}%
\usepackage{booktabs}%
\usepackage{algorithm}%
\usepackage{algorithmicx}%
\usepackage{algpseudocode}%
\usepackage{listings}%
\usepackage{adjustbox}
\usepackage{mathdots}

\theoremstyle{thmstyleone}%

\theoremstyle{thmstyletwo}%

\theoremstyle{thmstylethree}%
\begin{document}

\title[Article Title]{Nonlocal Microwave Engineering: Shaping Dispersion Relations and Enabling Frequency-Momentum Transformations via Time-Switched Long-Range Interactions}
\author[1]{\fnm{Matteo} \sur{Ciabattoni}}\email{mc2574@cornell.edu}

\author[1]{\fnm{Francesco} \sur{Monticone}}\email{francesco.monticone@cornell.edu}

\affil*[1]{\orgdiv{School of Electrical and Computer Engineering}, \orgname{Cornell University}, \orgaddress{\city{Ithaca}, \state{New York}, \postcode{14850}, \country{USA}}}

\abstract{Nonlocal metamaterials (MTMs) have recently attracted significant attention across different areas of wave physics, owing to their ability to translate long-range interactions among meta-atoms into a wide array of wavevector-dependent responses and functionalities. In this work, we introduce nonlocal transmission line metamaterials (TL MTMs) as a versatile platform to investigate and engineer nonlocality in the microwave frequency regime. We first establish a concise theoretical framework for nonlocal TL MTMs based on circuit and network theory, from which we derive the general dispersion relation for TL MTMs with arbitrarily complex nonlocal coupling configurations. Building upon this foundation, we demonstrate how such structures can be used to synthesize nearly arbitrary even-symmetric dispersion relations, effectively linking nonlocal circuit parameters to prescribed dispersion profiles. We then introduce time-switched nonlocal TL MTMs, a new class of metamaterials with time-varying nonlocality in which the nonlocal branches are dynamically activated as an electromagnetic pulse propagates through the structure. This platform enables complex, nearly arbitrary frequency-momentum transformations on a propagating pulse, as well as the excitation of modes with positive, negative, and zero group velocity. Finally, we experimentally validate our theoretical and numerical predictions with a proof-of-concept demonstration of a time-switched nonlocal TL MTM, observing a vertical transition in the dispersion diagram induced by abrupt time-switching. Our results provide new physical insights into the behavior of nonlocal MTMs, establish a versatile platform to investigate the interplay of frequency dispersion, spatial dispersion and time modulation, and, more broadly, lay a general foundation for the design of more advanced nonlocal and time-varying electromagnetic and photonic systems.
}

\maketitle
\section*{Introduction}\label{Intro}
Transmission line metamaterials (TL MTMs) have been extensively studied in both one and two dimensions \cite{caloz2005electromagnetic} and have been successfully exploited both to demonstrate a wide range of novel wave-physics phenomena, including negative refraction \cite{eleftheriades2002planar,lai2004composite,iyer2002negative} and subwavelength focusing \cite{sedighy2013wideband}, and for numerous practical microwave-engineering applications, such as the design of dual band couplers, conventionally restricted to a single band of operation \cite{lin2004arbitrary}, and directional couplers with arbitrary coupling levels and zero electrical length \cite{caloz2004novel}. Traditional TL MTMs are periodic structures formed by interconnecting lumped-circuit elements or transmission-line segments in a standard ladder network to realize desired effective properties or unusual frequency-wavevector relationships. However, TL MTMs have so far been limited to \emph{local} interactions or, in other words, periodic connections between neighboring meta-atoms. Networks with nearest-neighbor connections are the simplest to implement but can greatly restrict the range of available degrees of freedom, especially in terms of dispersion engineering as we shall see in the following. 

Recently, there has been growing interest in going beyond  such a local assumption, both in the study of the optical/electromagnetic response of natural materials, which has revealed naturally occurring nonlocal effects, and in the design of engineered MTMs \cite{chen2025nonlocal,monticone2025nonlocality} with strong artificial nonlocality. In such structures, effective nonlocality (also known as spatial dispersion, corresponding to a wavevector-dependent response) emerges from multipolar effects or long-range interactions mediated by guided modes or physical nonlocal connections. In the latter case, each meta-atom is not only connected to its nearest neighbor (as in conventional TL MTMs) but also to its $m$th long-distance neighbor through additional, periodically repeated nonlocal branches. This architecture has been studied across several areas of wave physics: in elastic systems using mass-spring chains \cite{chen2021roton,kazeomi2023drawing,paul2024complete}, in acoustics \cite{wang2022nonlocal,martinez2021experimental} using tubes as connections, as well as in electrodynamics with BNC, Bayonet Neill-Concelman, cables \cite{chen2023cable}. These studies have revealed a range of promising results, including the ability to support dispersion regions with opposite phase and group velocity — leading to negative refraction with virtually no loss \cite{martinez2021experimental}— wide zero-group velocity (or flat-band) regions \cite{chen2023cable}, and exotic roton and maxon-like dispersion features \cite{chen2021roton,kazeomi2023drawing} analogous to those observed in the acoustical response of superfluid Helium-4 \cite{landau1941theory,godfrin2021dispersion}. 

Perhaps the most exciting feature of this class of nonlocal MTMs is that each additional nonlocal connection contributes a term of the form $\cos(mka)$ to the dispersion relation \cite{chen2025nonlocal,kazeomi2023drawing,paul2024complete}, where 
$k$ is the wavenumber, $a$ is the physical size of the unit cell ($ka$ is its electrical length), and $m$ denotes the connection order between meta-atoms $n$ and $n \pm m$. Consequently, it becomes possible---at least in principle---to engineer the dispersion diagram of a nonlocal MTM to match any even Fourier series of a desired function within the first Brillouin zone. This concept has already been theoretically demonstrated in the elastic-wave domain \cite{kazeomi2023drawing,paul2024complete}, with the synthesis of broad classes of dispersion relations in one- and two-dimensional spring–mass chain models. However, as we will show, microwave circuit implementations provide additional degrees of freedom that allow for even finer control over the dispersion characteristics of the nonlocal MTM with fewer practical constraints.

In the following, we first derive a concise theoretical model for TL MTMs, establishing them as a versatile platform to study arbitrarily complex nonlocal, discrete, electromagnetic structures, uncovering new physical insights, opportunities, and limitations. We then demonstrate a powerful application of our theoretical framework: the synthesis of nearly arbitrary, even, dispersion functions for electromagnetic waves. We then extend this concept to time-switched nonlocal MTMs, showing that rapidly ``switching on'' nonlocal branches in time enables the selective excitation of specific modes, as well as the direct transition to intricate dispersion relations, corresponding to complex frequency-momentum transformations on a propagating pulse. Finally, we validate our theoretical and numerical results, for both the time-invariant and time-switched cases, by designing, fabricating, and experimentally characterizing a nonlocal TL MTM prototype.

\section*{Nonlocal transmission-line metamaterials theory}
The dispersion relation of TL MTMs can be derived in several ways, but perhaps the most practical approach is through the use of ABCD matrices \cite{collin2001foundations}, which relate the input and output currents and voltages as \([V_{\text{in}} \; I_{\text{in}}]^T = 
\begin{bmatrix} A & B \\ C & D \end{bmatrix}
[V_{\text{out}} \; I_{\text{out}}]^T\). The unit cell of a standard, local TL MTM is conventionally modeled as a shunt admittance $Y_l$ with a series impedance $Z_l/2$ on both sides \cite{caloz2005electromagnetic}. Using the cascading property of ABCD matrices, the overall matrix of the unit cell is given by the product of the individual matrices of the shunt and series elements:
\begin{equation}
    \begin{bmatrix}
        A & B \\
        C & D
    \end{bmatrix}_{\text{Unit Cell}}
    = 
    \begin{bmatrix}
        1 & Z_{l}/2\\
        0 & 1
    \end{bmatrix}
    \begin{bmatrix}
        1 & 0 \\
        Y_{l} & 1
    \end{bmatrix}
    \begin{bmatrix}
        1 & Z_{l}/2\\
        0 & 1
    \end{bmatrix}
    = \begin{bmatrix}
        \frac{Y_lZ_l}{2} + 1 & Z_l\left(\frac{Y_lZ_l}{4} + 1\right) \\
        Y_l & \frac{Y_lZ_l}{2} + 1
    \end{bmatrix}
    \label{eq1}
\end{equation}
To find the dispersion relation of the periodic network, we can then use the standard equation $\text{cosh}(ka) = \frac{A + D}{2}$ \cite{collin2001foundations}, where $A,D$ are the ABCD unit cell parameters. More information on the derivation of this equation is provided in Supplementary Material 1.
\begin{figure}
    \centering\includegraphics[width=1\linewidth]{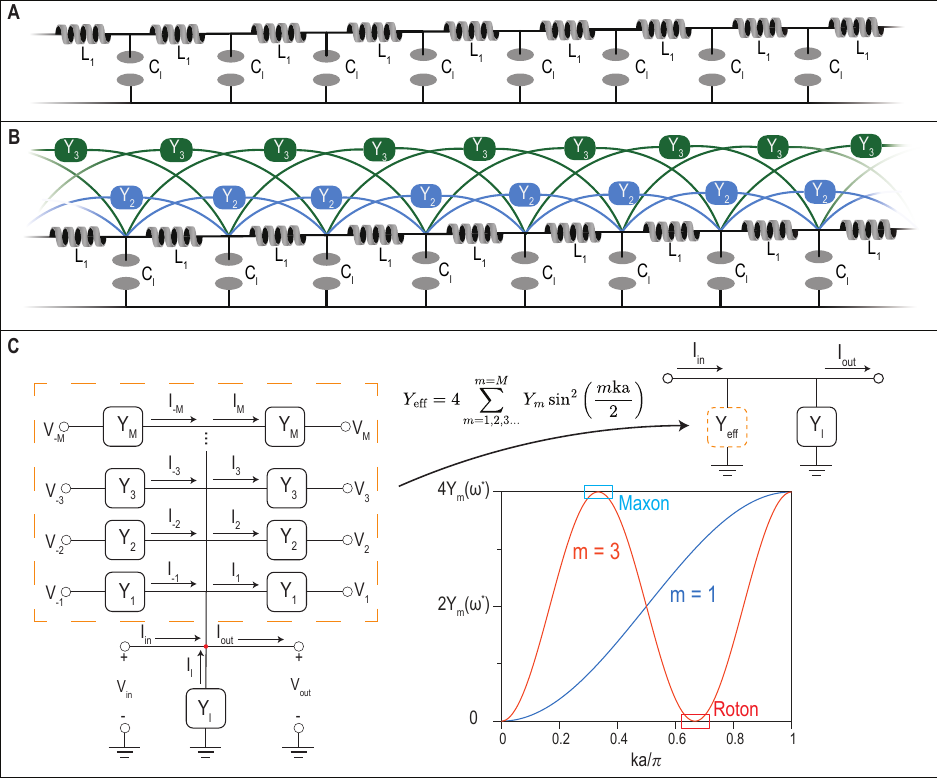}
    \caption{\textbf{Theory of microwave nonlocal transmission-line metamaterials (TL MTMs).} \textbf{(A)} Standard, local, LC Ladder network composed of shunt capacitors, $C_{l}$ each connected to their nearest neighbor through an inductor $L_{1}$.
    \textbf{(B)} Nonlocal TL MTM having local shunt components, $C_l$ connected to their nearest neighbor through an inductor $L_{1}$ and their $m =2,3$ longer-distance neighbors via arbitrary admittances $Y_{2}$, in blue, and $Y_{3}$ in green.
    \textbf{(C)} ``Super-cell'' of a nonlocal TL MTM: the local shunt admittance $Y_l$ is connected to $M$ long-distance neighbors through nonlocal admittances, $Y_m$. Using Kirchhoff’s current law, we can simplify the orange boxed part into an effective admittance given by a simple sum of nonlocal admittances $Y_m$ each modulated by the wavenumber. The equivalent unit cell of the nonlocal MTM is shown with the effective admittance on the top-right corner. The inset of the figure shows the variation of the modulated nonlocal admittance as a function of the wavenumber, at a fixed frequency $\omega_{*}$, within the first Brillouin Zone of the MTM. The modulated nonlocal admittance achieves a maximum value of $4Y_m$ (boxed in blue), which then corresponds to a ``maxon''  in its dispersion relation in Fig. \ref{fig2}(A), and a minimum value of 0 (boxed in red), and therefore an open circuit, which then corresponds to a ``roton'' in the dispersion relation.}
    \label{fig1}
\end{figure}

Using this approach, it is instructive to begin our study with what is arguably the simplest TL MTM structure: the LC ladder network pictured in Fig. \ref{fig1}(A). To find its dispersion relation, we substitute $Z_{l} = j\omega L_{1}$ and $Y_{l} = j\omega C_{l}$ into Eq. \ref{eq1}, where $L_{1}$, $C_{l}$, and $\omega$ are the local inductance, local capacitance, and angular frequency, respectively. 
After some algebra, this leads to the dispersion relation of a lumped-element transmission line:
\begin{equation}
    \omega(ka) = 2\omega_{c}\sqrt{\sin^{2}\left(\frac{ka}{2}\right)},
    \label{eq2}
\end{equation}
where $\omega_{c}=1/\sqrt{L_{1}C_{l}}$. Then, we introduce arbitrarily many nonlocal branches connecting shunt capacitors $n$ and $n \pm m$ via an admittance $Y_{m}$, as shown in Fig. \ref{fig1}(B) for $m = 2,3$. It is now more challenging to define a unit cell compared to the local LC ladder case since each capacitor now connects to its nearest neighbor (through $Z_{l}= j\omega L_{1}$)  as well as to arbitrarily many capacitors along the line (through the nonlocal admittances $Y_{m}$).  Furthermore, while a standard ABCD matrix relates the voltages and currents of a unit cell ($V_n$ and $I_n$) to those of its immediate neighbor ($V_{n+1}$ and $I_{n+1}$) \cite{collin2001foundations}, here the connections extend beyond nearest neighbors. Thus, it may appear that the ABCD matrix method cannot be used to derive the dispersion relation of such a nonlocal structure. However, we can greatly simplify the circuit by defining a preliminary “super-cell”, as shown in Fig. \ref{fig1}(C) and then applying Kirchhoff’s current law (KCL) at the central node (highlighted in red on the ``super-cell''):
\begin{multline*}
    I_{\text{in}} = V_{\text{in}}\left[Y_{l} + Y_{2}\left(2 - e^{j2ka} - e^{-j2ka}\right) + Y_{3}\left(2 - e^{j3ka} - e^{-j3ka}\right)\right. \\
    \left. + \cdots + Y_{M}\left(2 - e^{jMka} - e^{-jNka}\right) \right] + I_{\text{out}}.
\end{multline*}
%where $Y_{l}$ is the local shunt admittance and $Y_{m}$ is the nonlocal admittance connecting nodes $n$ and $n \pm m$. 
Using the identity $2 - e^{-jmka} - e^{jmka} = 4\sin^{2}\left(\frac{mka}{2}\right)$, we can simplify this equation to:
\begin{equation}
    I_{\text{in}} = V_{\text{in}}\left[
    Y_{l} 
    + 4Y_{2}\sin^{2}\left(ka\right)
    + 4Y_{3}\sin^{2}\left(\frac{3ka}{2}\right)
    + \cdots 
    + 4Y_{M}\sin^{2}\left(\frac{Mka}{2}\right)\right] 
    + I_{\text{out}}.
    \label{eq3}
\end{equation}
Finally, recalling the definition of an ABCD matrix (see above), and comparing it with Eq. \ref{eq3}, together with the fact that $V_{\text{in}} = V_{\text{out}}$ (also shown in Fig. \ref{fig1}(C)), allows us to obtain an ABCD matrix representation for the nonlocal branches:
\begin{equation}
    \begin{bmatrix}
        A & B \\
        C & D
    \end{bmatrix}_{\text{Effective Admittance}}
    = 
    \begin{bmatrix}
        1 & 0\\
        Y_{\mathrm{eff}}=4\sum_{m} Y_{m}\sin^{2}\left(\frac{mka}{2}\right) & 1
    \end{bmatrix}
    \label{eq4}    
\end{equation}
This means that we can model all the nonlocal branches as a single, effective admittance given by the sum of the nonlocal admittances $Y_m$ modulated by the wavenumber of the wave propagating inside the structure, where the period of the modulation is simply set by which nodes $Y_m$ connects.
This allows us to simplify the``super-cell'' of the nonlocal TL MTM to a unit cell that resembles that of a local TL MTM as illustrated in Fig. \ref{fig1}(C). 

It is interesting to focus for a moment on the implications of Eq. \ref{eq4}. First, the effective admittance $Y_{\mathrm{eff}}$ is a sum of squared sinusoids. Since each term is an even function of $ka$ reciprocity is automatically enforced within the structure (i.e., waves propagating in opposite directions experience the same properties). In the inset of Fig. \ref{fig1}(C), we plot the modulated nonlocal admittance as a function of wavenumber, for a fixed frequency $\omega_*$, for two orders of nonlocality, $m = 1$ (local case, in blue) and $m = 3$, (in orange) within the first Brillouin Zone ($ka \in [0,\pi]$). In the plot, the modulated nonlocal admittance reaches a maximum value of $4Y_{m}$ for waves having wavenumber $k = \frac{\pi}{3a}$, as well as a minimum value of $Y_{m} = 0$ for $k = \frac{2\pi}{3a}$. At this latter value of wavenumber, waves will not ``see'' the nonlocality since the nonlocal branch acts as an open circuit. Intuitively, to decrease the period of the sinusoid with respect to $ka$, adding more local maxima and minima within the first Brillouin zone, we must connect unit cells at longer distances, namely, increase the order of spatial nonlocality.

Having established a framework to derive an ABCD matrix representation of a nonlocal TL MTM, with an effective admittance modulated by the wavenumber, we can now compute the dispersion relation of any nonlocal structure of this type, beginning with the following: an LC ladder network with each shunt capacitor being connected to its immediate neighbor as well as its $m$th neighbor, with $m = 3$, through an inductor. We term the local inductor as $L_1$ and the nonlocal one as $L_3$. The dispersion relation of such a structure is:
\begin{equation}
        \omega(ka) = 2\omega_{c}\sqrt{\sin^{2}\left( \frac{ka}{2}\right) + \frac{L_{1}}{L_{3}}\sin^{2}\left(\frac{3ka}{2}\right)}
        \label{eq5}
\end{equation}

\begin{figure}
    %\centering
    \begin{adjustbox}{width=1.2\linewidth, center}
    \includegraphics[keepaspectratio]{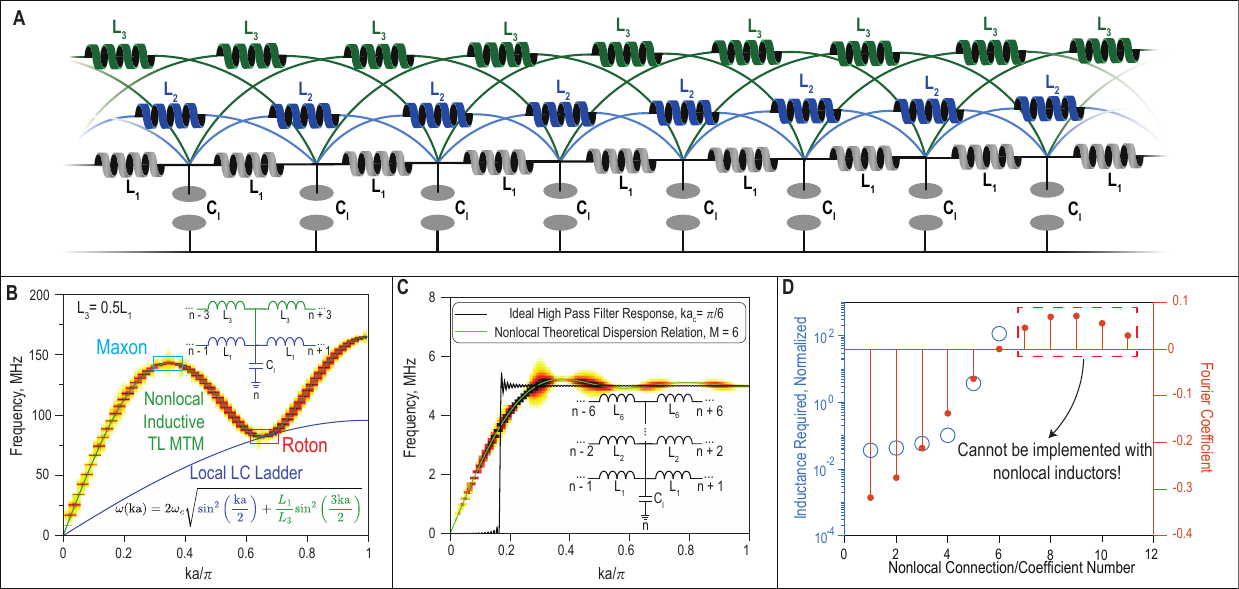}
    \end{adjustbox}
    \caption{\textbf{Inductive nonlocal TL MTM.} \textbf{(A)} Illustration of an inductive nonlocal TL MTM with two orders of nonlocality $m = 2,3$. Each shunt local capacitor is connected to its $1^{\text{st}}$(in gray), $2^{\text{nd}}$ (in blue) and $3^{\text{rd}}$ (in green) neighbor, with the last two representing nonlocal connections via an inductor having inductance $L_{m}$.
    \textbf{(B)} Dispersion relation of a nonlocal inductive TL MTM: theoretical (green line) and numerical (contour plot) results. A “super-cell” of the nonlocal TL MTM used to implement the dispersion relation is shown in the top-right inset of the panel, with the local components in blue and the nonlocal ones in green. For comparison, the dispersion relation of the local LC ladder is plotted in blue. The nonlocal dispersion is always above that of the local one due to the squared sinusoidal term introduced in the numerator of the expression, a feature of nonlocal TL MTM of inductive type. At the roton point shown in Fig. \ref{fig1}(C) the nonlocal dispersion coincides with the local one since the wave does not “see” the nonlocal inductance.
    \textbf{(C)} High-pass-filter-like (HPF-like) dispersion relation implemented with an inductive nonlocal TL MTM of order $M = 6$ (theoretical results in green, and contour plot for the numerical results). In black we show the ideal high-pass-filter-like dispersion relation implemented using an ideal Fourier Series. A “super-cell” of the nonlocal TL MTM is shown in the inset of the figure. Due to passivity, and therefore the inability to realize positive Fourier coefficients, this is the best approximation of the desired dispersion relation achievable using only nonlocal inductors.
    \textbf{(D)} Plot showing the Fourier coefficients (stem plot; right orange $y$-axis) needed to realize the desired dispersion relation pictured in (C) and the nonlocal inductance values (circles) required to implement such coefficients (left blue $y$-axis). As indicated by Eq. \ref{eq7}, only the first 6 Fourier coefficients, which are negative, can be implemented using passive nonlocal inductors. This strongly restricts the range of synthetizable functions using nonlocal TL MTM of inductive type.}
    \label{fig2}
\end{figure}

Comparing this equation with the the local case of Eq. \ref{eq2}, we see that it retains the local response as indicated by the presence of the $\sin^{2}\left( \frac{ka}{2}\right)$ term, but the nonlocal connection now introduces a new sinusoidal term $\sin^{2}\left(\frac{3ka}{2}\right)$, where the 3 comes from connecting each local shunt capacitor with their $m=3$ long-distance neighbor, as expected. Interestingly, the nonlocal sinusoidal term has a coefficient of $\frac{L_{1}}{L_{3}}$, namely, the ratio of local to nonlocal inductance, which determines the strength of the nonlocal connection, or, in other words, how much the presence of nonlocality affects the overall dispersion. Its form is very intuitive in this simple case: as $L_3$ gets smaller, the strength of the nonlocality grows since the nonlocal inductance shorts the local LC unit cells. The opposite occurs for larger values of $L_3$ for which the nonlocal elements approach open circuits and do not affect the local dispersion.

To validate our theoretical approach and insights, Fig. \ref{fig2}(B) shows the theoretical dispersion relation given by Eq. \ref{eq5} (green curve), for the case with $L_3 = 0.5 L_1$, together with the numerical dispersion relation obtained by solving the state-space equations of a finite inductive nonlocal TL MTM ($N = 300$), evaluating the spatio-temporal Fourier transform of each node voltage, and plotting the result as a contour plot. %\blue{in green by plotting Eq. \ref{eq5} and by solving the state-space equations of a finite inductive nonlocal TL MTM (N = 300), we plot the spatio-temporal Fourier transform of each node voltage as a contour plot.
The agreement between numerical and theoretical results is excellent. Note that the contour plot has a non-zero width due to the finite number of nodes considered in the simulation.
(For additional details on the numerical experiments, refer to Supplementary Section 3.) As can be seen in the figure, the introduction of the nonlocal inductor generates regions of negative dispersion and saddle points. The plot reaches a maximum point, also termed in the acoustic/mechanical literature as a ``maxon'' \cite{kazeomi2023drawing,chen2021roton,zhang2024observation}, at the same location in which the modulated nonlocal admittance reaches its maximum (shown in Fig. \ref{fig1}(C)). The dispersion also reaches a local minimum, the ``roton'', when the nonlocal admittance vanishes and the corresponding branch becomes an open circuit; predictably, the nonlocal dispersion coincides with the local one at this point, since the wave does not “see” the presence of nonlocality. Both of these points, the roton and the maxon, are critical points of the dispersion relation with zero-group velocity, occurring at non-trivial values of wavenumber/momentum, and are analogous to the Van Hove singularities \cite{van1953occurrence} already observed in electronic band structures.
Comparing the admittance plot in Fig. \ref{fig1}(C) with Fig. \ref{fig2}(B), we see that the region of negative group velocity between the maxon and the roton coincides with the region where the modulated nonlocal admittance decreases as a function of increasing wavenumber (and, therefore, the effective nonlocal inductance increases in value, all the way to an open circuit).
This explains the onset of negative dispersion, as the dispersion relation must connect the maxon and roton points, whose separation is characterized by increasing wavenumber and \emph{decreasing} frequency (due to the modulated nonlocal admittance decreasing to zero). %Consequently, the dispersion relation must connect the maxon and roton points, whose separation is dictated by an increasing wavenumber and decreasing frequency, giving rise to a region of negative dispersion.
This is analogous to the creation of saddle points and negative group velocity regions in optical plasmonic-dielectric stratified systems \cite{karalis2005surface,karalis2009plasmonic} (we will further discuss the relation between these systems and our nonlocal platform in the Conclusion).
We also note that this unusual dispersion relation has already been demonstrated with acoustic/elastic waves  \cite{martinez2021experimental,kazeomi2023drawing,paul2024complete}, based on mass-spring chains where the nonlocal connections consist of springs of different spring constants. 

\subsection*{Constructing dispersion relations with nonlocal metamaterials}

Having studied what is arguably the simplest nonlocal TL MTM, we can now turn to a more complex case in which arbitrarily many nonlocal inductive connections are introduced. This scenario is illustrated in Fig. \ref{fig2}(A) for a representative example with two nonlocal connections ($m = 2,3$) for illustrative purposes. The dispersion relation becomes:
\begin{equation}
    \omega^{2} = \sum_{m = 1,2,\dots}^{m=M} \frac{2}{L_{m}C_{l}}\left[1 - \cos(mka)\right],
    \label{eq6}
\end{equation}
where we used the trigonometric identity $4\sin^{2}\left(\frac{x}{2}\right) = 2 - 2\cos(x)$ (cf. Eq. \ref{eq5}).
Then, considering that the Fourier series of an even periodic function can be written as
$ f(x) = a_0 + \sum_m a_m \cos(mx)$, and comparing the dispersion relation in Eq. \ref{eq6} with this Fourier series, we can map each nonlocal component into a Fourier coefficient : 
\begin{equation*}
    a_{0} =  \sum_{m}\frac{2}{C_{l}L_{m}}
\end{equation*}
\begin{equation}
    a_{m} = -\frac{2}{L_{m}C_{l}}
    \label{eq7}
\end{equation}
This is a very intriguing mapping that, as already noted in \cite{kazeomi2023drawing} for the acoustic/mechanical case, allows one to exploit nonlocal connections to synthesize unusual dispersion diagrams. There are, however, two major restrictions inherent to this nonlocal TL MTM that a Fourier series does not have. First, by examining Eq. \ref{eq6}, we see that $\omega(k = 0) = 0$. This is due to the inductive nature of our nonlocal TL MTM: signals at low frequencies will incur very short phase delays traveling through the structure due to the series inductors along the line, which operate very closely to short-circuits in this low-frequency regime. The second, and more stringent constraint, is due to passivity, which implies that the lumped-element inductances and capacitances in our network must be positive (otherwise the stored electric and magnetic energy, or the potential/kinetic energy in the mechanical case, would be negative). This, in turn, implies that in the mapping above we cannot implement arbitrary signed coefficients, but only negative ones, as indicated by Eq. \ref{eq7}. Thus, passivity strongly restricts our ability to design arbitrary, reciprocal dispersion relations using inductive nonlocality (or spring-based nonlocality in the mechanical case) since a Fourier series takes advantage of coefficients of both signs to implement most periodic functions. %\red{We also briefly note that, in addition to passivity, relativistic causality also restrict the dispersion functions that can be implemented, as further discussed later.}

To illustrate the impact of these restrictions, in Fig. \ref{fig2}(C) we attempt to realize a dispersion diagram that approximates the Fourier series of a high-pass filter (HPF) response (plotted in black) within the first Brillouin Zone of an inductive nonlocal TL MTM. In Fig. \ref{fig2}(D), we plot the Fourier coefficients of such a series (on the right $y$-axis) together with the value of nonlocal inductance (on the left $y$-axis) needed to implement such coefficients. For $m\leq6$, since the coefficients are negative, they can be implemented with nonlocal inductors. The resulting dispersion relation, for a nonlocal TL MTM of order $M=6$, is shown in Fig. \ref{fig2}(C), calculated both analytically (green) and numerically (contour) using the approach described above. The dispersion diagram qualitatively follows a high-pass-like response, but it is far from a good approximation of the desired function. This result, however, cannot be improved by introducing more nonlocal connections, and therefore more Fourier terms, because the next Fourier coefficients are positive (Fig. \ref{fig2}(D)), whereas Eq. \ref{eq7} shows that, by only using passive components, the realized coefficients, $a_{m}$, are always negative for an inductive nonlocal TL MTM. Despite their limitations highlighted here, inductive nonlocal TL MTMs can still be used to generate other, unusual, and potentially useful, dispersion relations, including the rational design of dispersion diagrams featuring several saddle points, as shown in Supplementary Section 4.

To overcome the major constraints of inductive nonlocal TL MTMs, we can leverage a second component not yet used in a nonlocal fashion: the capacitor. Interestingly, in spring-mass nonlocal acoustic/mechanical systems, capacitive nonlocality would correspond to a nonlocal mass connection, rather than a spring connection. This could potentially be realized with a more exotic component called an ``inerter''\cite{chen2009missing}, which however has not yet been demonstrated in this context. The fact that, in our case, this type of nonlocality can be realized with a simple capacitor highlights the potential and flexibility of electromagnetic systems to explore nonlocal wave physics phenomena. To illustrate the behavior of a capacitive nonlocal TL MTM, we consider a single nonlocal connection implemented through a capacitance $C_3$ linking nodes $n$ and $n\pm3$ and derive its dispersion relation using the same ABCD-matrix approach used for the inductive case, obtaining
\begin{equation}
    \omega(ka) = 2\omega_{c}\sqrt{\frac{\sin^{2}\left(\frac{ka}{2}\right)}{1 + 4\frac{C_3}{C_l}\sin^{2}\left(\frac{3ka}{2}\right)}}.
    \label{eq8}
\end{equation}

\begin{figure}
    \centering
    \includegraphics[width=1\linewidth]{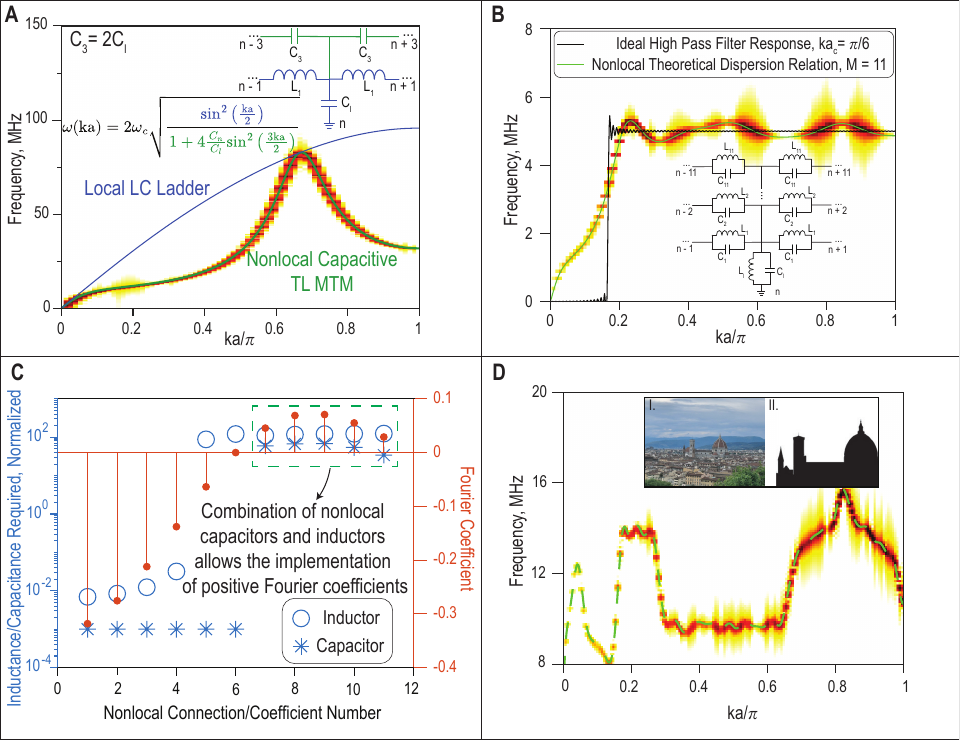}
    \caption{\textbf{Constructing more complex dispersion curves with capacitive and inductive nonlocality.} \textbf{(A)} Dispersion relation of a nonlocal capacitive TL MTM: theoretical (green line) and numerical (contour plot) results. . A “super-cell” of the nonlocal capacitive TL MTM used to implement the dispersion relation is shown in the top-right inset of the panel, with the local components in blue and the nonlocal ones in green. For comparison, the dispersion relation of the local LC ladder is shown in blue. The nonlocal dispersion relation always lies below that of the local case because the sinusoidal term appears in the denominator of the expression, in contrast to the inductive case, where it appears in the numerator. \textbf{(B)} High-pass-filter-like dispersion relation implemented with a capacitive nonlocal TL MTM of order $M$ = 11 (theoretical results in green, and contour plot for the numerical results). In black we show the ideal HPF-like dispersion relation implemented using an ideal Fourier Series. A “super-cell” of the nonlocal TL MTM is shown in the inset of the figure: we utilize a parallel LC resonator both in the shunt central component, as well as for local and nonlocal connections. \textbf{(C)} Plot showing the Fourier coefficients (stem plot; right orange y-axis) needed to realize the desired dispersion relation pictured in (B) and the nonlocal inductance values (circles) and capacitance values (asterisks) required to implement such coefficients (left blue y-axis).  The first 6 coefficients, which are negative, can be implemented with inductors only. For $m > 6$, instead, both nonlocal inductors and capacitors are needed. As discussed in the main text, the introduction of nonlocal capacitors allows for the implementation of positive Fourier coefficients and, therefore, the realization of a broader range of periodic dispersion functions. \textbf{(D)} Nonlocal TL MTM whose dispersion relation follows the profile of the Florence-Duomo. The first Brillouin Zone of the MTM is shown with theoretical (green dashed line) and numerical (contour plot) results. In the top-left inset (I), we show a picture of the Florence Duomo view from Piazzale Michelangelo and in the top-middle inset (II) we show the silhouette of the Duomo used to design the MTM. The MTM “super-cell” is identical to that shown in the inset of (B), but here with $M = 50$. This highlights how higher orders of nonlocality enable the realization of increasingly complex dispersion features, albeit at the cost of increased structural size.}
    \label{fig3}
\end{figure}

\noindent It is interesting to compare the dispersion relation of nonlocal TL MTMs of inductive and capacitive type, given by Eq. \ref{eq5} and Eq. \ref{eq8}, respectively. While both expressions retain the local sinusoidal term $\sin^{2}\left(\frac{ka}{2}\right)$ and introduce a nonlocal squared sinusoidal term, in the inductive case the nonlocal term is in the numerator, whereas it appears in the denominator in the capacitive case. The reason for this can be traced to the difference in phase response between an inductor and a capacitor with respect to an applied voltage. Furthermore, in both cases, we can express the strength of nonlocality with a ratio of the local to nonlocal component (the ratio is flipped in the capacitive case). We plot the dispersion relation of the capacitive nonlocal structure in Fig. \ref{fig3}(A) for $C_3 = 2C_l$. As can be seen, this dispersion relation also exhibits critical points and regions of negative dispersion due to the introduction of the nonlocal capacitance; however, in this case the dispersion lies below the local dispersion relation (shown in blue), in contrast to the inductive case, where it lies above it (Fig. \ref{fig2}(B)). Similar to the inductive case, the dispersion relation of the nonlocal structure coincides with that of the local one at $k = \frac{2\pi}{3a}$, when the nonlocal admittance becomes an open circuit. Due to passivity, capacitive nonlocal TL MTM suffer from similar general limitations as inductive ones.

Having established the two main building blocks, and canonical examples, of nonlocal TL MTMs---namely, networks with purely inductive and purely capacitive nonlocality---we can now derive the dispersion relation for a more general class of nonlocal TL MTM. Specifically, we consider structures in which the shunt admittance, as well as the local and nonlocal connections, are parallel LC resonators (the series LC resonator case can be studied in a similar way). A ``super-cell'' of such a structure is shown in the inset of Fig. \ref{fig3}(B). The resulting dispersion relation is:
\begin{equation}
        \omega^{2} = \frac{-1/L_{l} + \sum_{m = 1,2,\dots}^{m = M}2/L_{m}\left[\cos(mka) - 1\right]}{- C_{l} + \sum_{m = 1,2,\dots}^{m = M}2C_{m}\left[\cos(mka) - 1\right]}
        \label{eq9}
\end{equation}
By comparing this dispersion relation to that of purely inductive and capacitive nonlocal networks, in Eq. \ref{eq6} and \ref{eq8}, respectively, we see that it retains key features of both: the inductive nonlocality, $L_m$, introduces a Fourier series in the numerator while the capacitive nonlocality, $C_m$, introduces one in the denominator. The addition of the nonlocal capacitance $C_{m}$, together with the local shunt inductance, $L_{l}$, relaxes both limitations of the inductive nonlocal TL MTM. In particular, examining Eq. \ref{eq9} shows that the shunt inductor placed parallel with the local capacitance $C_l$ allows the dispersion relation to have a non-zero $\omega$-intercept, i.e., $\omega(k=0) \neq 0$, introducing a lower-frequency cutoff for the propagating mode. This occurs because a local TL MTM with shunt parallel LC resonators behave as a band-pass filter, preventing propagation below one of its resonance frequencies. Additional details on the propagation characteristics of such a local TL MTM are provided in Supplementary Materials Section 2.
More importantly, introducing a nonlocal capacitance on each nonlocal branch provides additional degrees of freedom in the design, enabling us to implement positive Fourier coefficients while retaining passivity and, therefore, positivity of the lumped-element capacitances and inductances. In other words, the Fourier series of the ratio of Fourier series on the right-hand-side of Eq. \ref{eq9} has signed coefficients even though the coefficients of the individual series are all negative (except $a_0$), which allows us to approximate the signed coefficients of the Fourier series of a desired dispersion function. This is demonstrated in Fig. \ref{fig3}(B) and (C), where we implement the same HPF-like dispersion relation of Fig. \ref{fig2}(C) and (D), now using our more general class of nonlocal TL MTMs. As seen from the coefficient plot in Fig. \ref{fig3}(C), the nonlocal capacitance is unnecessary for the first six Fourier coefficients of the desired function, which are negative and can therefore be implemented purely with nonlocal inductors (note that the capacitance values are very small, approaching an open circuit, for $m < 6$). However, once the Fourier coefficients become positive, a combination of inductors and capacitors allows us to implement these positive coefficients and, therefore, more closely approximate the desired HPF-like dispersion relation. As another example, we can also implement the opposite response, namely, a low-pass-filter-like dispersion relation, featuring negative dispersion over a wide bandwidth, as shown in Supplementary Section 5. These examples show that, by employing parallel LC resonators as both local and nonlocal connections, we drastically increase the ``expressivity'' of the nonlocal network to approximate positive, even, periodic dispersion functions described by a Fourier series inside the first Brillouin zone. While a rigorous proof of whether the right-hand-side of Eq. \ref{eq9} can truly approximate \emph{any} physical dispersion function of this type goes beyond the scope of this paper, our numerical experiments are particularly promising.

To further demonstrate this point, we design a nonlocal TL MTM whose dispersion relation follows the profile of the Florence Duomo as shown in Fig. \ref{fig3}(D). This is done by first extracting the silhouette of the Duomo, shown in Fig. \ref{fig3}(D)(II) and converting its shape and features into a one-dimensional target profile. Then, a TL MTM with a high nonlocal order is optimized to approximate this profile as its dispersion relation by leveraging $M = 50$ nonlocal connections. In the main plot of Fig. \ref{fig3}(D), we show the theoretical dispersion relation obtained by evaluating Eq. \ref{eq9} (green dashed curve) and the numerical dispersion relation determined by solving the circuit's state-space equations (contour plot). As the figure shows, the agreement between theoretical and numerical results is excellent, demonstrating the rather striking ability of nonlocal TL MTM to synthesize extremely complex dispersion functions with a wide range of unusual features. 

We also note that we have verified that the slope of this dispersion band, corresponding to the group velocity in this lossless scenario, remains below the free-space speed of light, $c$, everywhere. %(the light line, shown in blue, is steeper, \red{approximately,} than all dispersion features). 
In general, issues of relativistic causality need to be considered carefully in the modeling of nonlocal systems, as the nonlocal connections may appear to provide ``shortcuts'' for a wave propagating along the structure, potentially leading to superluminal group velocity if the wave in the local TL already propagates close to $c$. Our model can be slightly modified to address this issue by adding suitable delays (e.g., in the form of TL segments) in the nonlocal branches to account for the actual propagation delay along these paths. In our case in Fig. \ref{fig3}(D), however, this is not an issue because we operate at sufficiently low frequencies that such propagation delays are minimal and can be safely neglected, greatly simplifying the analysis and design of the dispersion relations of nonlocal TL MTMs (in contrast, propagation delays becomes important at higher frequencies, with the effect of smoothing the dispersion features such that the local slope remains below $c$ everywhere).  

\begin{figure}
    \centering    \includegraphics[width=0.95\linewidth,keepaspectratio]{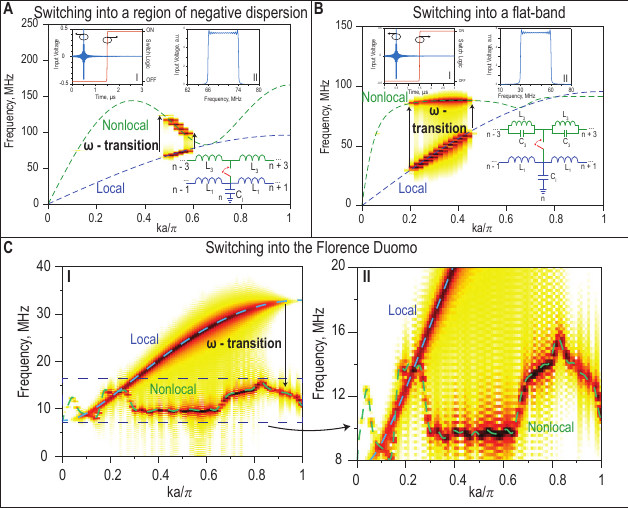}
    \caption{\textbf{Time-switched nonlocal TL MTM.} \textbf{(A)} Local LC transmission line excited with a sinc pulse of bandwidth 66-74 MHz (shown in time domain in inset I, and in frequency domain in inset II). The spatio-temporal Fourier transform of the propagating pulse coincides with the local dispersion relation (dashed blue line) given by Eq. \ref{eq2}. While the pulse is fully contained within the LC ladder, the nonlocal inductive branch is switched on as shown in inset I (orange line shows the switching waveform). The pulse undergoes an $\omega-$transition from a region of positive to a region of negative group velocity. The spatio-temporal Fourier transform of the pulse after switching agrees falls exactly on the dispersion curve of the nonlocal structure, given by Eq. \ref{eq5} for $L_3$ = $L_1$ (dashed green line). The “super-cell” of the time-switched nonlocal TL MTM is shown in the bottom right corner, with the nonlocal branch in green and the local one in blue. In red we show the switch that can dynamically connect and disconnect the nonlocal branch from the TL MTM. \textbf{(B)} Similar to panel (A) but for a sinc pulse of bandwidth 30-60 MHz and a time-switching-induced $\omega-$transition into a flat-band region of the dispersion curve of the nonlocal structure. % Local LC transmission line excited with a sinc pulse of bandwidth 30-60 MHz (shown in time domain in inset I, and in frequency domain in inset II). The spatial Fourier transform of the pulse coincides exactly with the theoretical local dispersion relation (dashed blue line) equation shown in Eq. \ref{eq1}. As the pulse is fully contained within the LC ladder, the nonlocal branch is switched on as shown in inset I. 
    To implement a large flat band, the nonlocal branch is designed with a parallel LC with: $L_3 = 0.3L_1$ and $C_3 = C_l$. %The pulse undergoes an $\omega-$transition from a region of positive group velocity, to zero group velocity due to the introduction of the nonlocal branch. The spatial Fourier transform of the pulse after switching agrees perfectly with the theoretical prediction of Eq. \ref{eq9} (dashed green line) for a single nonlocal connection. The “super-cell” of the time-switched nonlocal TL MTM is shown in the bottom right corner, with the nonlocal branch in green and the local one in blue. 
    \textbf{(C)} Local LC transmission line excited with a sinc pulse of bandwidth 5-40 MHz. While the pulse is fully contained within the LC ladder, $M = 50$ nonlocal connections are switched on. In this case, the $\omega-$transition shapes the pulse to follow the dispersion curve approximating the profile of the Florence Duomo (I). A close-up view is shown in (II). In both the local and nonlocal case, the agreement between theoretical predictions (blue and green dashed lines) and numerical experiments (contour plots) is excellent. These results highlight the ability of time-switched nonlocal TL MTMs to shape propagating modes on demand. Time-domain field animations for the cases in panels (A) and (B) are provided as Supplementary Materials.}
    \label{fig4}
\end{figure}

Another important observation to make concerns the relationship between the complexity of the dispersion curve and the physical size of the structure that implements it. As demonstrated throughout this work, increasing the order of nonlocality, i.e., the number of nonlocal connections, allows for the synthesis of an increasing number of Fourier coefficients, thereby enabling more intricate dispersion relations. Inevitably, this comes at the cost of greater structural complexity and increased physical size, since sufficient space is required to accommodate the nonlocal connections while avoiding overlap. This observation highlights a direct parallel with the concept of overlapping nonlocality introduced in Ref. \cite{miller2023why}, which identifies the number of overlapping nonlocal input-output interactions (corresponding to the number of independent channels that must cross a transverse aperture) as a fundamental reason for the need for thickness in optics and wave physics. Here, nonlocal TL MTMs embody a version of this fundamental trade-off: increasing the complexity of the desired dispersion curve requires a larger order of nonlocality and, therefore, a larger physical size. The Florence Duomo dispersion relation exemplifies this: while nonlocal TL MTMs can realize elaborate dispersion curves featuring sharp details and regions of positive, negative and zero group velocity, achieving such behaviors requires $M = 50$ nonlocal connections. % - thereby significantly increasing both the structural complexity and the physical size of the system.  
We speculate that, as in Ref. \cite{miller2023why} for input-output transfer functions, there exist strict fundamental limitations on the physical size of any structure---not just the nonlocal, discrete, electromagnetic structures studied here, but any wave-physics system---designed to realize a prescribed dispersion curve. Formalizing this intriguing observation will be the subject of future work.

\subsection*{Time-switching in nonlocal metamaterials}
Finally, we extend the concept of nonlocal TL MTMs by breaking the assumption of time invariance, introducing a new class of metamaterials with time-varying nonlocality that highlights the potential of combining nonlocal effects with temporal modulation. Specifically, we focus on time-switched nonlocal TL MTMs, in which we introduce a switch between the local and nonlocal branches that enables the nonlocal coupling to be abruptly activated or deactivated.  
It is well established that, when a structure experiences an abrupt temporal change, such as a sudden modification of its characteristic impedance, while a pulse is propagating through it, the wave's frequency and energy content are altered, whereas its momentum is conserved if the temporal change is spatially homogeneous \cite{morgenthaler2003velocity,fante2003transmission}. This has been experimentally demonstrated earlier for water waves \cite{bacot2016time} and, more recently, for electromagnetic waves in a local TL MTM \cite{moussa2023observation}.  
%If however, the structure lacked spatial translational symmetry the time-switched event would cause a $\omega-k$ translation \cite{ciabattoni2025observation}.

Here, we apply the concept of temporal switching to three illustrative examples of nonlocal TL MTMs to demonstrate how controlled frequency transitions can be used to excite modes with unique dispersion properties simply by switching on the nonlocal branches, corresponding to drastic frequency–momentum transformations imparted on a propagating pulse. In our first example, Fig. \ref{fig4}(A), a local LC ladder is excited with a sinc pulse with a bandwidth, 66 - 74 MHz (as shown in \ref{fig4}(A)(II)), chosen to populate the desired range of wavevectors on the blue dashed line (local dispersion relation). Once the pulse is fully contained within the structure, the nonlocal branch (which in this case is an inductor, $L_3 = L_1$, connecting nodes $n$ and $n \pm 3$, as shown in the inset of \ref{fig4}(A)) is rapidly switched on. The temporal switching event induces a vertical frequency transition, as indicated in Fig. \ref{fig4}(A), transforming the propagating mode from one with positive group velocity to one with negative group velocity. 
In our second example, we follow a similar approach, but now with the goal to induce a ``frozen'' pulse within the structure. In this case, we design a nonlocal TL MTM featuring a parallel LC resonator in the nonlocal branch connecting nodes $n$ and $n \pm 3$, as illustrated in Fig \ref{fig4}(B). When the nonlocality is switched on, the initially forward-propagating mode is transformed into a frozen mode with zero group velocity by transitioning into the flat-band region of the nonlocal dispersion relation, as demonstrated in Fig. \ref{fig4}(B), thereby effectively trapping the pulse inside the structure. %This is very well shown in Fig. \ref{fig4}(B) in which the mode transitions from having positive group velocity to a flatband. 
This effect is clearly illustrated in Section 4 of the Supplementary Materials, where time-domain field animations show the temporal evolution of the pulse as it abruptly comes to rest within the MTM. 
We also note that, to plot the numerical results in Fig. \ref{fig4}, we take the Fourier transform of the nodal voltages for the entire simulation time (before and after the switching event), which explains why our contour plots show both modes, before and after switching. In reality, the transition is complete as shown in the time-domain field animations in Supplementary Materials. The contour plots also display colored regions that do not correspond to either mode, but instead appear between them. These features arise because the switching event is not modeled as perfectly abrupt, but rather as exponential (the switching waveform is shown in \ref{fig4}(A)(I) and \ref{fig4}(B)(I) in orange); this is done to avoid numerical instabilities, shorten simulation time, and more closely model a realistic scenario. For more information on the numerical methods, see Supplementary Materials Section 3.

Finally, we demonstrate a more extreme example of temporal nonlocality by employing the nonlocal Florence-Duomo TL MTM design introduced in Fig. \ref{fig3}(D). A sinc pulse is launched into the corresponding local TL MTM, exciting a mode with smooth dispersion and positive group velocity. When $M = 50$ nonlocal connections are abruptly activated, this simple propagating mode is reshaped into one whose dispersion mirrors the complex Florence Duomo profile, as shown in Fig.  \ref{fig4}(C). This example highlights the versatility and power of nonlocal TL MTMs: having already established that their dispersion can be engineered in a nearly arbitrary manner, these results show that, by introducing temporal switching, one can impart nearly arbitrary frequency-momentum transformations on a propagating pulse by transforming an otherwise ordinary medium (such as an LC-ladder TL MTM) into one that exhibits intricate dispersion characteristics.

\subsection*{Experimental proof-of-concept demonstration}
\begin{figure}
    \centering
  \includegraphics[width=1\linewidth,keepaspectratio]{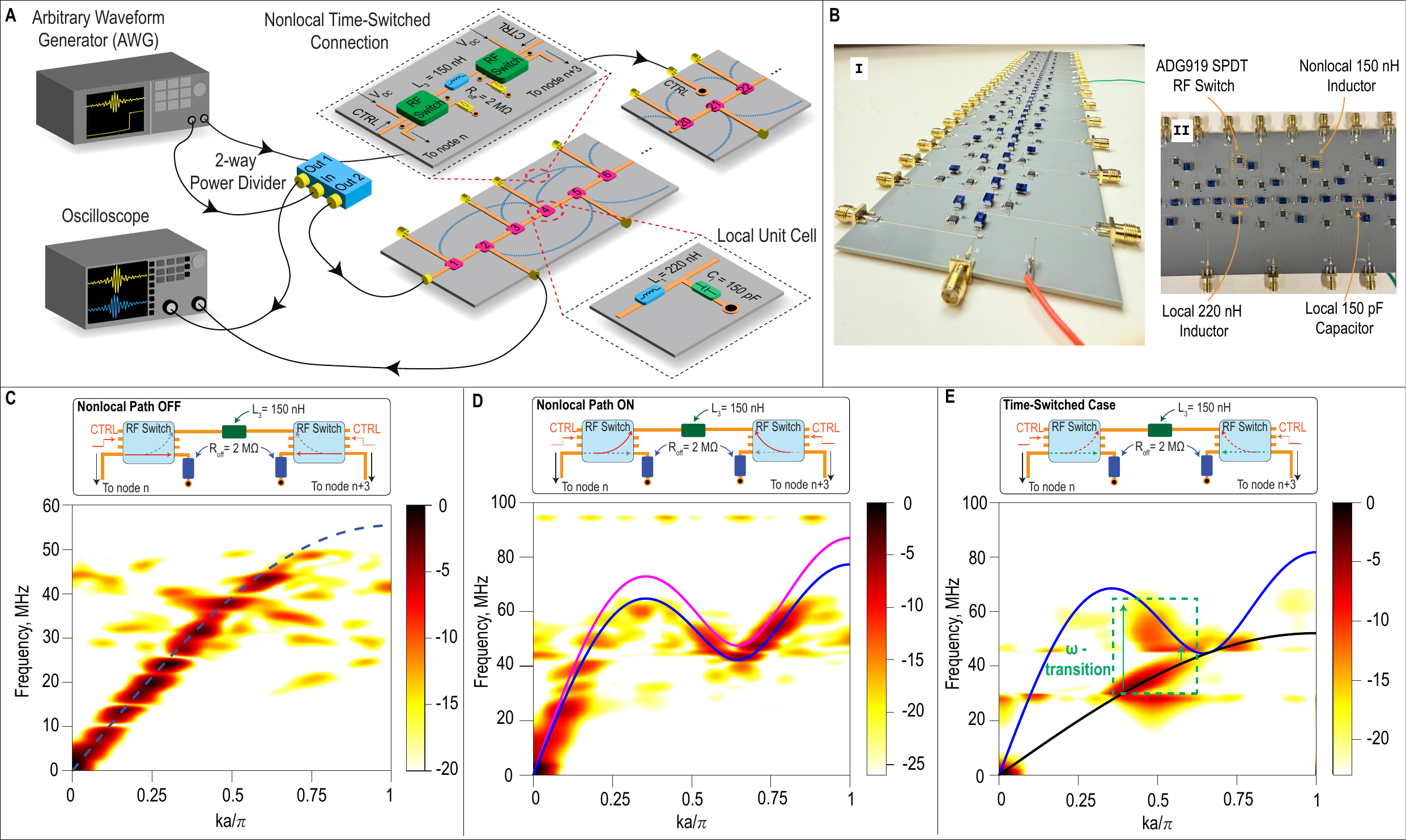}
    \caption{\textbf{Experimental Demonstration of Time-Switched Nonlocal TL MTM}. 
    \textbf{(A)} Measurement setup used to extract the experimental dispersion relation of the time-switched TL~MTM. An arbitrary waveform generator (AWG) provides both the input excitation and control signals, while an oscilloscope records the voltage at each node. The TL~MTM consists of unit cells comprising a 220~nH series inductor and a 150 pF shunt capacitor. Each shunt capacitor is coupled to its $n\pm3$ neighbor through a time-switched nonlocal connection, implemented using two SPDT microwave switches that alternate between a 2 M$\Omega$ shunt resistor and a 150 nH nonlocal inductor. \textbf{(B)} Photographs of the nonlocal TL MTM. 
    \textbf{(C)}Experimental dispersion relation (contour plot in dB scale) of the TL MTM with the nonlocal path switched OFF, as shown in the inset. The black curve indicates the theoretical dispersion relation of the local TL MTM. The structure is excited with a sinc pulse having a maximum frequency of 95 MHz.  
    \textbf{(D)} Same experiment as in (C) but with the nonlocal path switched ON. The pink curve shows the theoretical dispersion relation computed with Eq. \ref{eq5}, whereas the blue curve is the dispersion relation computed with the same equation but accounting for a higher local capacitance originating from the microstrip and stripline connections from the local capacitor to the nonlocal inductor.
    \textbf{(E)} Experimental dispersion relation (contour plot in dB scale) of the nonlocal TL MTM when the nonlocal path is switched ON as the pulse is propagating through the structure. Because the pulse initially travels along the local TL MTM, its spectral content initially follows the local dispersion curve (black curve). Abrupt activation of the nonlocal path while the pulse resides inside the structure produces a vertical transition, transferring energy to the higher-frequency nonlocal dispersion curve (blue curve). The excitation is a sinc pulse with a bandwidth of 30--45 MHz.}
    \label{fig5}
\end{figure}

To validate our theoretical predictions and numerical simulations for both time-invariant and time-switched nonlocal TL MTMs, we design and fabricate a nonlocal TL MTM on a printed circuit board (PCB) incorporating RF switches. The structure consists of a local LC ladder composed of 42 nodes of series inductors (220 nH) and shunt capacitors (150 pF), as illustrated in the unit cell schematic in Fig. \ref{fig5}(A). Each local capacitor is connected to its $n\pm3$ neighbor through a time-switchable nonlocal coupling implemented using two single-pole double-throw (SPDT) RF switches per connection. The first output of a switch terminates in a shunt 2 M$\Omega$ resistor to ground and the second output connects to the other switch via a 150 nH inductor, as shown in Fig. \ref{fig5}(A) in the ``non-local time-switched connection'' panel. This configuration enables toggling between local and nonlocal operation by adjusting the control voltage. Pictures of the PCB are shown in Fig. \ref{fig5}(B). For more information on the circuit design and fabrication, please refer to the Methods section.

Experimental results for the time-invariant case are presented in Fig. \ref{fig5}(C) and (D). To obtain the local dispersion relation, the control voltage is set to $V_{\text{low}}$, directing each RF switch to connect the local capacitors to the shunt resistors, which effectively behave as open circuits. A broadband sinc pulse (bandwidth 0 - 90 MHz) is then injected, and the node voltages are measured across all 42 nodes of the TL MTM, yielding the local dispersion relation through a spatio-temporal Fourier transform. Additional details of the experimental set-up are provided in the Methods section. When the same experiment is repeated with the control voltage set to $V_{\text{high}}$, the structure instead exhibits the behavior of an inductive nonlocal TL MTM. Experimental results for the local (Fig. \ref{fig5}(C)) and nonlocal (Fig. \ref{fig5}(D)) cases, are in close agreement with the theoretical predictions of Eq. \ref{eq2} and \ref{eq5}, respectively. Note that, in the nonlocal case, we plot two theoretical dispersion relations (blue and pink), both computed using Eq. \ref{eq5}. The blue curve uses an increased local capacitance to account for the additional capacitance introduced by the microstrip and strip-line connections from the local shunt capacitor to the nonlocal inductor.

The results for the time-switched system are shown in Fig.  \ref{fig5}(E). A narrowband sinc pulse (bandwidth of 30-45 MHz) is launched with the control voltage initially set to $V_{\text{low}}$, allowing the pulse to propagate along the local TL MTM. Once the pulse reaches the midpoint of the structure, all RF switches are abruptly actuated, activating the nonlocal coupling paths. As shown in Fig. \ref{fig5}(E), this produces the expected vertical transition in the frequency domain, i.e., a frequency up-conversion of the pulse toward the higher-frequency dispersion curve of the nonlocal TL MTM. Our experimental results are influenced by several non-idealities. Most importantly, as mentioned above, the nonlocal inductors are connected to the local capacitors through microstrip and stripline segments rather than ideal links, introducing delays and parasitic inductance and capacitance. Each line also includes two microstrip-to-stripline via transitions, adding further parasitics that modify the dispersion relation.
Further discussion of additional non-idealities impacting the results, together with the data-processing procedure used here, is provided in the Methods section.
Despite these non-idealities, which could be mitigated in future integrated implementations, the experimental results match the numerically predicted transfer of energy toward the higher-frequency nonlocal dispersion curve shown in Fig. \ref{fig4}(A).

\section*{Conclusion}
In conclusion, we have introduced and investigated nonlocal TL MTMs in the microwave frequency regime through theory, numerical simulations, and laboratory experiments. Our new theoretical approach shows that, regardless of the number of nonlocal connections, the unit cell of a nonlocal TL MTM can be greatly simplified by modeling nonlocal branches as a single effective admittance modulated by the wavenumber. This representation significantly facilitates the analysis and design of such structures. Using this theoretical framework, we have studied two canonical examples of nonlocal TL MTMs, namely, networks with purely inductive and purely capacitive nonlocality, unveiling intriguing design opportunities as well as practical and fundamental challenges due to passivity. Then, by combining these two building blocks, we have demonstrated the potential of a broader class of nonlocal TL MTMs to synthesize intricate, nearly arbitrary dispersion relations within their first Brillouin Zone, as illustrated by examples such as high-pass-filter-like or Florence-Duomo-like dispersion relations. We then introduced time-switched nonlocal TL MTMs, in which the dynamic activation of nonlocal branches of the periodic structure can induce drastic dispersion changes and enable on-demand nearly arbitrary frequency-momentum transformations on a propagating pulse. Finally, we have experimentally validated our theoretical predictions by fabricating and characterizing a prototype of this first-of-its-kind nonlocal, time-varying, TL MTMs, demonstrating strong agreement between measured and theoretical results.

Looking ahead, our results may open new research opportunities for nonlocal wave-physics systems, establishing nonlocal TL MTM as a versatile platform for studying and exploiting complex nonlocal effects. Future research efforts could focus on possible extensions to two-dimensional TL MTM platforms and to more complex time-varying systems, as well as on developing a more rigorous understanding of fundamental tradeoffs between dispersion complexity, order of nonlocality, and physical size. We also anticipate that these findings may influence other areas of nonlocal wave physics, ranging from acoustical and mechanical systems—where capacitive-like nonlocality remains unexplored—to optical systems, where our approach may inspire new, systematic design strategies for realizing on-demand dispersion relations. In particular, at optical frequencies it would be interesting to explore a combination of the concepts introduced here with those developed in Refs. \cite{karalis2009plasmonic,karalis2005surface}, where stacks of transversely homogeneous plasmonic and dielectric layers were used to create unusual dispersion features, such as negative dispersion and saddle points. Although the physical platforms are different, we believe that the same underlying physics of nonlocal, long-range interactions (mediated, in Refs. \cite{karalis2009plasmonic,karalis2005surface}, by guided waves along different layers) is at play in both cases. This connection suggests intriguing opportunities to engineer nearly arbitrary dispersion relations and frequency–wavevector transformations at optical frequencies by translating some of the ideas proposed here to the optical domain, potentially in the form of stacks of nonlocal metasurfaces \cite{shastri2023nonlocal}.

\section*{Methods}

\bmhead{Printed Circuit Board Design}
Fig. \ref{fig5}(A) illustrates the local unit cell, which comprises a Coilcraft HA4033-ALC 220 nH ceramic-core inductor and a Kyocera AVX KGM15ACG1H151JT 150 pF capacitor. A large-footprint inductor (SMD size 1812) with a ceramic core was selected to minimize losses and dispersion at high-MHz frequencies. Each local unit cell is connected to two reflective single-pole double-throw (SPDT) microwave switches (Analog Devices ADG919). The two outputs of each switch are terminated by a 2~M$\Omega$ resistor (Vishay-Dale MCT0603MD2004BP500) shunted to ground and a Coilcraft HA4031-ALC 150 nH inductor, respectively; the latter provides the connection from the local unit cell to its $n \pm 3$ neighbor. Since the RF switch has two states, we can toggle the nonlocal path ON or OFF: when the CTRL voltage is set to $V_{\text{high}}$, each local capacitor is connected via the switches and the nonlocal inductance to its $n \pm 3$ neighbors. However, when the CTRL voltage is set to $V_{\text{low}}$, each local capacitor is connected in shunt to a 2 M$\Omega$ resistor, thereby turning OFF the nonlocal path.

 All inductors are chosen to be ceramic core to achieve high Q factor and self-resonance frequency. Despite this, both local and nonlocal inductors have a modest DCR of approximately 100 $\text{m}\Omega$, which inevitably renders pulse propagation lossy throghout the structure and makes it challenging to excite  the edge of the Brillouin Zone, as shown in Fig. \ref{fig5} (C) and (D).
Because all inductors have ceramic cores, their magnetic fields are not fully self-contained and can couple to nearby local or nonlocal inductors. To minimize this interference, each inductor is spaced at least 8 mm from its neighbors (approximately 1.5 times body length); where this clearance cannot be maintained, adjacent inductors are oriented perpendicular to one another.
 
The interleaving of local and nonlocal connections, as also shown in Fig. \ref{fig2}(A), necessitates careful routing to avoid path crossings and maintain signal isolation. To this end, we employ an eight-layer PCB fabricated on FR-4 ($\varepsilon_{r} = 4.6$, $\delta = 0.03$ at 1~GHz) with outer and inner copper weights of 1~oz and 0.5~oz, respectively. All components are mounted on the top layer, and local connections are routed as microstrip traces on that same layer. Nonlocal connections, along with control and DC signal routing, are implemented as stripline traces on the inner layers.
 
The final prototype consists of 42 nodes and measures $56.4 \text{ cm} \times 8.8 \text{ cm}$

\bmhead{Measurement Set-up}
The measurement setup is illustrated in detail in Fig. \ref{fig5}(A). A Tektronix AWG520 arbitrary waveform generator produces both the input signal and the control step function. The input pulse is defined by:

\begin{equation*}
    V_{\text{in}} = \text{sinc}\left(2\pi f_{\text{max}}t\right) - \frac{f_{\text{min}}}{f_{\text{max}}}\text{sinc}\left(2\pi f_{\text{min}}t\right)
\end{equation*}
 For experiments in which the nonlocal path remains in a fixed state (either ON or OFF) throughout the measurement, $f_{\text{min}}$ and $f_{\text{max}}$ are set to 0 and 95 MHz. In the time-switched experiment, where only a narrow portion of the Brillouin zone of the local TL MTM is of interest, $f_{\text{min}}$ and $f_{\text{max}}$ are set to 30 and 45 MHz, respectively.

The input pulse is routed through a Mini-Circuits ZMSC-2-1+ two-way power divider. One output is connected to an Infiniivision DSOX2022A oscilloscope to record the input waveform and synchronize the AWG with the oscilloscope; the other output is connected to the DUT (device under test).

To obtain the experimental dispersion relation of the DUT, the voltage at each node must be measured. For this purpose, each local capacitor is connected to the edge of the PCB via a short transmission line (maximum length of 33 mm, corresponding to an electrical length of approximately $5^{\circ}$ at the highest recorded frequency). The oscilloscope is then connected sequentially to each node to record all 42 nodal voltages. To minimize the noise collected in each measurement, we use the oscilloscope's averaging option to measure average 256 samples before collecting the data.

The control step function, with $V_{\text{low}} = 0$ V and $V_{\text{high}} = 1.5$ V, is applied at the center of the board (near unit cell number 21, as shown in Fig. \ref{fig5}(A)) to minimize the switching delay between switches located at opposite ends of the board.
Finally, the switches are powered by a DC bias of 2 V supplied by an Agilent E3630A power supply (not shown in Fig. \ref{fig5}(A)).
\bmhead{Data Analysis}
Once all nodal voltages have been recorded, they are assembled into a matrix in which each column corresponds to a node and each row corresponds to a discrete time step. A Fourier transform is then applied along the time axis to obtain the frequency spectrum of each nodal voltage. Because the measured data can be noisy, particularly in the time-switched case, a normalization procedure is applied to improve the clarity of the dispersion features. At each frequency bin, the root-mean-square (RMS) magnitude is computed across all nodes, yielding a single scalar per frequency that characterizes the overall signal level at that frequency. A noise floor is then defined as 10\% of the peak RMS value across all frequencies, and the spectral data are normalized by dividing each frequency bin by the greater of its spatial RMS value or the noise floor. This prevents artificial amplification of frequency components where the signal level is low relative to the noise.
We then take the Fourier transform of the frequency domain data to compute the dispersion relation. This data is normalized to the maximum value recorded and plotted in dB scale.

\backmatter

\bmhead{Supplementary information}
Supplementary Sections 1–6.

\bmhead{Acknowledgments}
F.M. thanks Prof. David A. B. Miller of Stanford University for helpful discussions during PQE 2025 on thickness limits in nonlocal systems. F.M. also thanks Prof. Steven Johnson of MIT and Prof. Pai Wang of University of Utah for bringing Refs. \cite{karalis2009plasmonic,karalis2005surface} and Refs. \cite{kazeomi2023drawing,paul2024complete,zhang2024observation}, respectively, to our attention.

\section*{Declarations}

\bmhead{Funding}
This work was supported by the Air Force Office of Scientific Research with grant no. FA9550-22-1-0204 and the Office of Naval Research with grant no. N00014-22-1-2486.

\bmhead{Conflict of interest/Competing interests}
The authors declare no competing interests.

\bmhead{Availability of data and materials}
Authors can confirm that all relevant data are included in the paper and/or its Supplementary Information files.

\bmhead{Code availability}
The code used to produce these results is available upon request from the corresponding author.

\bmhead{Authors' contributions} F.M. conceived the idea and supervised the project. M.C. performed the theoretical calculations, the circuit design for each nonlocal TL MTM example, designed the numerical experiment and led the writing of the manuscript with contributions from all authors. All authors participated in the conceptual development, discussion of the results, and the writing of the manuscript.

\bibliography{sn-bibliography}

@article{paul2024complete,
  title = {Complete inverse design to customize two-dimensional dispersion relation via nonlocal phononic crystals},
  author = {Paul, Sharat and Hasan, Md Nahid and Fu, Henry Chien and Wang, Pai},
  journal = {Phys. Rev. B},
  volume = {110},
  issue = {14},
  pages = {144304},
  numpages = {10},
  year = {2024},
  month = {Oct},
  publisher = {American Physical Society},
  doi = {10.1103/PhysRevB.110.144304},
  url = {https://link.aps.org/doi/10.1103/PhysRevB.110.144304}
}

@article{kazeomi2023drawing,
  title = {Drawing Dispersion Curves: Band Structure Customization via Nonlocal Phononic Crystals},
  author = {Kazemi, Arash and Deshmukh, Kshiteej J. and Chen, Fei and Liu, Yunya and Deng, Bolei and Fu, Henry Chien and Wang, Pai},
  journal = {Phys. Rev. Lett.},
  volume = {131},
  issue = {17},
  pages = {176101},
  numpages = {7},
  year = {2023},
  month = {Oct},
  publisher = {American Physical Society},
  doi = {10.1103/PhysRevLett.131.176101},
  url = {https://link.aps.org/doi/10.1103/PhysRevLett.131.176101}
}

@article{chen2021roton,
    author = {Chen, Yi and Kadic, Muamer and Wegener, Martin},
    title = {Roton-like acoustical dispersion relations in 3D metamaterials} ,
    journal = {Nature Communications},
    year = {2021},
    doi = {10.1038/s41467-021-23574-2},
    volume = {12},
    issue = {1},
    pages = {3278}
}

@book{caloz2005electromagnetic,
  title={Electromagnetic metamaterials: transmission line theory and microwave applications},
  author={Caloz, Christophe and Itoh, Tatsuo},
  year={2005},
  publisher={John Wiley \& Sons},
    address= {Hoboken, NJ},
}

@book{collin2001foundations,
  title={Foundations for Microwave Engineering},
  author={Collin,Robert E.},
  year={2001},
  publisher={Wiley-IEEE Press},
  address={Piscataway, NJ},
}

@article{chen2023cable,
    author = {Chen, Yi and Abouelatta, Mahmoud A. A. and Wang, Ke and Kadic, Muamer and Wegener, Martin},
    title = {Nonlocal Cable-Network Metamaterials},
    journal = {Advanced Materials},
    volume = {35},
    number = {15},
    pages = {2209988},
    year = {2023}
}

@article{chen2025nonlocal,
    author ={Chen, Yi and Fleury, Romain and Seppecher, Pierre and Hu, Gengkai and Wegener, Martin} ,
    title = {Nonlocal metamaterials and metasurfaces},
    journal = {Nature Reviews Physics},
    year = {2025}
}

@article{miller2023why,
  title    = "Why optics needs thickness",
  author   = "Miller, David A B",
  journal  = "Science",
  volume   =  379,
  number   =  6627,
  pages    = "41--45",
  month    =  jan,
  year     =  2023,
  language = "en"
}

@article{wang2022nonlocal,
  title     = "Nonlocal interaction engineering of {2D} roton-like dispersion
               relations in acoustic and mechanical metamaterials",
  author    = "Wang, Ke and Chen, Yi and Kadic, Muamer and Wang, Changguo and
               Wegener, Martin",
  journal   = "Commun. Mater.",
  publisher = "Springer Science and Business Media LLC",
  volume    =  3,
  number    =  1,
  month     =  may,
  year      =  2022,
  language  = "en"
}

@article{martinez2021experimental,
author = {Julio Andrés Iglesias Martínez  and Michael Fidelis Groß  and Yi Chen  and Tobias Frenzel  and Vincent Laude  and Muamer Kadic  and Martin Wegener },
title = {Experimental observation of roton-like dispersion relations in metamaterials},
journal = {Science Advances},
volume = {7},
number = {49},
pages = {eabm2189},
year = {2021},
}

@article{landau1941theory,
  title = {Theory of the Superfluidity of Helium II},
  author = {Landau, L.},
  journal = {Phys. Rev.},
  volume = {60},
  issue = {4},
  pages = {356--358},
  numpages = {0},
  year = {1941},
  month = {Aug},
  publisher = {American Physical Society},
}

@article{godfrin2021dispersion,
  title = {Dispersion relation of Landau elementary excitations and thermodynamic properties of superfluid $^{4}\mathrm{He}$},
  author = {Godfrin, H. and Beauvois, K. and Sultan, A. and Krotscheck, E. and Dawidowski, J. and F\aa{}k, B. and Ollivier, J.},
  journal = {Phys. Rev. B},
  volume = {103},
  issue = {10},
  pages = {104516},
  numpages = {33},
  year = {2021},
  month = {Mar},
  publisher = {American Physical Society},
}

@article{eleftheriades2002planar,
  author={Eleftheriades, G.V. and Iyer, A.K. and Kremer, P.C.},
  journal={IEEE Transactions on Microwave Theory and Techniques}, 
  title={Planar negative refractive index media using periodically L-C loaded transmission lines}, 
  year={2002},
  volume={50},
  number={12},
  pages={2702-2712}
}

@article{lai2004composite,
  author={Lai, A. and Itoh, T. and Caloz, C.},
  journal={IEEE Microwave Magazine}, 
  title={Composite right/left-handed transmission line metamaterials}, 
  year={2004},
  volume={5},
  number={3},
  pages={34-50},}

@article{van1953occurrence,
  title={The occurrence of singularities in the elastic frequency distribution of a crystal},
  author={Van Hove, L{\'e}on},
  journal={Physical Review},
  volume={89},
  number={6},
  pages={1189},
  year={1953},
  publisher={APS}
}

@inproceedings{iyer2002negative,
  title={Negative refractive index metamaterials supporting 2-D waves},
  author={Iyer, Ashwin K and Eleftheriades, George V},
  booktitle={2002 IEEE MTT-S International Microwave Symposium Digest (Cat. No. 02CH37278)},
  volume={2},
  pages={1067--1070},
  year={2002},
  organization={IEEE}
}

@article{moussa2023observation,
  title={Observation of temporal reflection and broadband frequency translation at photonic time interfaces},
  author={Moussa, Hady and Xu, Gengyu and Yin, Shixiong and Galiffi, Emanuele and Ra’di, Younes and Al{\`u}, Andrea},
  journal={Nature Physics},
  volume={19},
  number={6},
  pages={863--868},
  year={2023},
  publisher={Nature Publishing Group UK London}
}

@article{fante2003transmission,
  title={Transmission of electromagnetic waves into time-varying media},
  author={Fante, R},
  journal={IEEE Transactions on Antennas and Propagation},
  volume={19},
  number={3},
  pages={417--424},
  year={2003},
  publisher={IEEE}
}

@article{morgenthaler2003velocity,
  title={Velocity modulation of electromagnetic waves},
  author={Morgenthaler, Frederic R},
  journal={IRE Transactions on microwave theory and techniques},
  volume={6},
  number={2},
  pages={167--172},
  year={2003},
  publisher={IEEE}
}

@article{bacot2016time,
  title={Time reversal and holography with spacetime transformations},
  author={Bacot, Vincent and Labousse, Matthieu and Eddi, Antonin and Fink, Mathias and Fort, Emmanuel},
  journal={Nature Physics},
  volume={12},
  number={10},
  pages={972--977},
  year={2016},
  publisher={Nature Publishing Group UK London}
}

@article{zhang2024observation,
  title={Observation of maxon-like ultrasound in elastic metabeam},
  author={Zhang, Peng and Liu, Yunya and Zhang, Keping and Wu, Yuning and Chen, Fei and Chen, Yi and Wang, Pai and Zhu, Xuan},
  journal={APL Materials},
  volume={12},
  number={3},
  year={2024},
  publisher={AIP Publishing}
}

@article{karalis2009plasmonic,
  title={Plasmonic-Dielectric Systems for High-Order Dispersionless Slow or Stopped Subwavelength Light},
  author={Karalis, Aristeidis and Joannopoulos, JD and Solja{\v{c}}i{\'c}, Marin},
  journal={Physical review letters},
  volume={103},
  number={4},
  pages={043906},
  year={2009},
  publisher={APS}
}

@article{lin2004arbitrary,
  title={Arbitrary dual-band components using composite right/left-handed transmission lines},
  author={Lin, I-Hsiang and DeVincentis, Marc and Caloz, Christophe and Itoh, Tatsuo},
  journal={IEEE Transactions on Microwave Theory and Techniques},
  volume={52},
  number={4},
  pages={1142--1149},
  year={2004},
  publisher={IEEE}
}

@article{caloz2004novel,
  title={A novel composite right-/left-handed coupled-line directional coupler with arbitrary coupling level and broad bandwidth},
  author={Caloz, Christophe and Sanada, Atsushi and Itoh, Tatsuo},
  journal={IEEE Transactions on Microwave Theory and Techniques},
  volume={52},
  number={3},
  pages={980--992},
  year={2004},
  publisher={IEEE}
}

@article{sedighy2013wideband,
  title={Wideband planar transmission line hyperbolic metamaterial for subwavelength focusing and resolution},
  author={Sedighy, S Hassan and Guclu, Caner and Campione, Salvatore and Amirhosseini, M Khalaj and Capolino, Filippo},
  journal={IEEE transactions on microwave theory and techniques},
  volume={61},
  number={12},
  pages={4110--4117},
  year={2013},
  publisher={IEEE}
}

@article{karalis2005surface,
  title={Surface-plasmon-assisted guiding of broadband slow and subwavelength light in air},
  author={Karalis, Aristeidis and Lidorikis, E and Ibanescu, Mihai and Joannopoulos, JD and Solja{\v{c}}i{\'c}, Marin},
  journal={Physical review letters},
  volume={95},
  number={6},
  pages={063901},
  year={2005},
  publisher={APS}
}

@article{monticone2025nonlocality,
  title={Nonlocality in photonic materials and metamaterials: roadmap},
  author={Monticone, Francesco and Mortensen, N Asger and Fern{\'a}ndez-Dom{\'\i}nguez, Antonio I and Luo, Yu and Zheng, Xuezhi and Tserkezis, Christos and Khurgin, Jacob B and Shahbazyan, Tigran V and Chaves, Andr{\'e} J and Peres, <? pag$\backslash$break?> Nuno MR and others},
  journal={Optical Materials Express},
  volume={15},
  number={7},
  pages={1544--1709},
  year={2025},
  publisher={Optica Publishing Group}
}

@article{chen2009missing,
  title={The missing mechanical circuit element},
  author={Chen, Michael ZQ and Papageorgiou, Christos and Scheibe, Frank and Wang, Fu-Cheng and Smith, Malcolm C},
  journal={IEEE circuits and systems magazine},
  volume={9},
  number={1},
  pages={10--26},
  year={2009},
  publisher={IEEE}
}

@article{shastri2023nonlocal,
  title={Nonlocal flat optics},
  author={Shastri, Kunal and Monticone, Francesco},
  journal={Nature Photonics},
  volume={17},
  number={1},
  pages={36--47},
  year={2023},
  publisher={Nature Publishing Group UK London}
}

\end{document}